\documentclass[11pt]{article}
\usepackage{epsfig,graphics}

\newcommand{\cerenkov}{\v{C}erenkov}
\newcommand{\Kstar}{\overline{K}^{\;\!*0}}
\newcommand{\eg}{e.g.}

\newcommand{\etal}{{\em et al.}}
\def\Journal#1#2#3#4{{#1} {\bf #2} (#4) #3}

\def\NIMA{{Nucl. Instrum. Meth.}}

\def\PLB{{Phys. Lett.}}
\def\PRL{Phys. Rev. Lett.}
\def\PRD{{Phys. Rev.}}

\def\CJP{Chin. J. Phys. (Taipei)}
\def\MPLA{{Mod. Phys. Lett.}}
\def\EPJC{{Eur. Phys. J.}}
\def\EPJD{{EPJdirect}}
\def\CPC{Comp. Phys. Comm.}

\def\APPB{{Acta Phys. Polon.}}

\def\issue(#1,#2,#3){{\bf #1} (#3) #2} 
\def\PRLX(#1,#2,#3){Phys.\ Rev.\ Lett.\ \issue(#1,#2,#3)}
\def\PLX(#1,#2,#3){Phys.\ Lett.\ \issue(#1,#2,#3)}
\def\PRDX(#1,#2,#3){Phys.\ Rev.\  \issue(#1,#2,#3)}
\def\NIMX(#1,#2,#3){Nucl.\ Instrum.\ Meth.\ \issue(#1,#2,#3)}
\def\IEEETNSX(#1,#2,#3){IEEE Trans.\ Nucl.\ Sci.\ \issue(#1,#2,#3)}
\def\CIP(#1,#2,#3){Comp.\ Phys.\ \issue(#1,#2,#3)}

\def\be{\begin{equation}}
\def\ee{\end{equation}}
\def\bea{\begin{eqnarray}}
\def\eea{\end{eqnarray}}

\begin{document}

\begin{table}[t]
\begin{center}
\begin{tabular}{lcl} 
2nd Frontiers in Contemporary Physics:  
& \hspace{15mm}   &  FERMILAB-Conf-01/236-E  \\ 

The Inner Space Outer Space Connection (FCP01) & &  UMS/HEP/2001-030 \\ 
5--10 Mar 2001, Nashville, Tennessee, USA & & \\
\end{tabular}
\end{center}
\end{table}
\vspace*{8mm}
\centerline{\LARGE \bf Search for Rare Charm Meson Decays at FNAL E791}
\vskip 16pt
\centerline{\Large D.~J.~Summers}
\vskip 3pt
\centerline{\large Department of Physics and Astronomy}
\vskip 3pt
\centerline{\large University of Mississippi-Oxford, University, MS 38677, USA}
\vskip 3pt
\centerline{\large e-mail: summers@relativity.phy.olemiss.edu}
\vskip 5pt
\centerline{\large (representing the Fermilab E791 Collaboration)}
\vskip 20pt
\leftline{\bf Abstract}
\vskip 3pt
We report the results of a {\it blind} 
search for flavor-changing neutral current (FCNC),
lepton-flavor violating, and lepton-number violating decays of $D^+$, 
$D_s^+$, and $D^0$
mesons (and their antiparticles) into 2--, 3--, and 4--body states including 
a lepton pair.
Such decays may involve Flavor-Changing Neutral Currents, Leptoquarks,
Horizontal Gauge Bosons, or Majorana Neutrinos.
No evidence for any of these decays is found.
Therefore, we present 90{\%} confidence level branching-fraction upper limits,
typically at the $10^{-4}$ level.
A total of 51 decay channels have been examined; 26 have not been previously
reported and 18 are significant improvements over previous results.

\vskip 20pt
\leftline{\bf Introduction}
\vskip 3pt
The E791 Collaboration has reported limits on rare 
and forbidden dilepton decays of charmed $D$ mesons 
\cite{FCNC,FCNCnew,4prong}. Such measurements probe the SU(2)$\times $U(1) 
Standard Model of electroweak interactions in search of new mediators 
and couplings \cite{Pakvasa,SCHWARTZ93}
beyond the $W^{\pm}$ and $Z^0$ discovered
at CERN in 1983 \cite{palmer}. 
Here we summarize the results 
of two related analyses. First \cite{FCNCnew} we examined the 
$\pi \ell \ell$ and $K\ell \ell$ decay modes of $D^+$ and $D_{s}^{+}$ 
and the $\ell^+ \ell^-$ decay modes of $D^0$. Then we extended the 
methodology to 27 dilepton decay modes of the $D^0$ meson \cite{4prong} 
containing either resonant $V\ell^+\ell^-$ decays, where $V$ is a 
$\rho ^0$, $\Kstar$, or $\phi$, and non-resonant $h_{1}h_{2}\ell \ell $ 
decays, where $h_{i}$ is either a $\pi $ or a $K$. The leptons were either 
muons or electrons. Charge-conjugate modes are implied
throughout this paper. The modes are lepton 
flavor-violating (\eg, \ $D^{+}\rightarrow \pi ^{+}\mu ^{+}e^{-}$), or 
lepton number-violating 
(\eg, \ $D_{s}^{+}\rightarrow \pi ^{-}\mu ^{+}\mu ^{+}$), or 
flavor-changing neutral current decays 
(\eg, \ $D^{0}\rightarrow \Kstar e^{+}e^{-}$). $WW$ box diagrams 
can mimic 
FCNC decays, but only at the $10^{-10}$ to $10^{-9}$ 
level \cite{SCHWARTZ93,Fajfer}. Long range effects through resonant 
modes (\eg, \ $D^{0}\rightarrow \Kstar \rho ^{0}, \ 
\rho^{0}\rightarrow  e^{+}e^{-}$) can occur at the $10^{-6}$ 
level \cite{Fajfer,Singer}. 
Neither Lorentz nor gauge
invariance \cite{CPT}
require lepton
number conservation.
A leptoquark \cite{LQ} in an exchange diagram (\eg, $D^0 \to \mu^+ e^-$) 
or a Horizontal Gauge Boson \cite{HGB} in a spectator diagram 
(\eg, $D^{+}\rightarrow \pi ^{+}\mu ^{+}e^{-}$)
might mediate a change both in quark and lepton
generation simultaneously. A Majorana Neutrino might lead to same sign dilepton 
decays \cite{Ali}.
Numerous experiments have studied rare 
decays of charge -1/3 strange quarks. Charge 2/3 charm quarks are 
interesting because they might couple differently \cite{Castro}.

The data come from measurements made with the E791 
spectrometer \cite{e791spect} at Fermilab's Tagged Photon Lab. 
The spectrometer has been upgraded for a series of charm experiments
including E516 \cite{e516}, E691 \cite{e691}, E769 \cite{e769}, and E791.
E791 events 
were produced by a 500 GeV/$c$~ $\pi ^{-}$ beam interacting in a fixed 
target consisting of five thin disks, one platinum and four diamond. 
The discs were well separated to allow charmed hadrons to decay 
in air spaces where other interactions would be minimal.
In addition to searching for rare decays,
E791 has searched for
$D^0 \, \overline{D}^{\,0}$ mixing \cite{mix}, {\it CP} 
violation \cite{CP}, and the pentaquark \cite{pentaquark}.
We have observed doubly cabibbo suppressed decays \cite{cabibbo}, discovered
that the $D_s^+$ has a longer lifetime than the $D^0$ \cite{Ds_life},
tagged quarks as being $c$ or $\overline{c}$
using $D$-$\pi$ production correlations \cite{Dpi}, and measured the 
$\Sigma^0_c \, - \, \Sigma^+_c$ mass splitting \cite{Sigma}.
Leading effects (e.g. more forward
$\pi^+(u\overline{d}) \rightarrow D^+(c\overline{d})$ than
$\pi^+(u\overline{d}) \rightarrow D^-(\overline{c}d)$) have been
observed in $D^+$ \cite{Dplus_prod} and $\Lambda_c^+$ 
\cite{Lambda_c_prod} production
and not in $D_s^+$ production \cite{Ds_prod}.  The total forward 
and differential cross
section for $D^0$ production has been measured \cite{D0_cross}. 
The differential cross section for $D^{*+}$ hadroproduction
has been measured \cite{dstar}. 
The decays 
$\Lambda_c^+ \to p K^- \pi^+$ \cite{Lambda_c}, 
$D^0 \to K^- K^+ \pi^- \pi^+$ \cite{KKpipi}, 
$D^0 \to K^- K^- K^+ \pi^+$ \cite{Sokoloff}, 
$D^+ \to K^- \pi^+ \pi^+$ \cite{Gobel}, and
$D_s^+ \to \pi^- \pi^+ \pi^+$ \cite{Ds3pi}
have been studied.  Evidence for the scalar meson $\sigma$(500)
in $D^+ \to \pi^- \pi^+ \pi^+$ decays has been observed \cite{D3pi}. 
In non-charm physics, E791 has measured 
$\Lambda^0$, $\Xi^-$, and $\Omega^-$ hyperon production asymmetries 
\cite{hyperon} and has used di-jet events to observe the pion valence quark
distribution \cite{Valence} and color transparency \cite{Color}.

E791 recorded a total of $2 \times 10^{10}$ events 
with a loose transverse energy requirement
and data acquisition system writing at 10 MB/s to a great wall of
42 8mm Exabyte tape drives \cite{da791}.
The resulting 50 Terabyte data set was totally unprecedented. It was
nevertheless reconstructed at parallel computing farms built for this purpose
at the University of Mississippi, Kansas State University, Fermilab, and
CBPF--Rio de Janeiro \cite{farm791} and yielded an unprecedented 200\,000 fully
reconstructed charmed hadron decays. 
Track and vertex
information came from ``hits'' in 23 silicon microstrip planes and 45 
wire chamber planes. This information and the bending provided by two 
dipole magnets were used for momentum analysis of charged particles. 
Kaon identification was carried out by two multi-cell \cerenkov{} 
counters \cite{Bartlett} that provided $\pi /K$ separation in the 
momentum range $6-60$~GeV/$c$. We required that the momentum-dependent 
light yield in the \cerenkov{} counters be consistent for kaon-candidate 
tracks, except for those in $D^0 \to \phi \pi^+ \pi^-$
decays with $\phi \rightarrow  K^{+}K^{-}$, 
where the narrow mass window for the $\phi$ decay provided sufficient 
kaon identification (ID). 

Electron ID was based on transverse shower shape plus matching wire 
chamber tracks to shower positions and energies in 
an electromagnetic calorimeter \cite{SLIC}.
The electron ID efficiency varied from 62$\%$ below 9 GeV/$c$ to 45$\%$ 
above 20 GeV/$c$. The probability to misidentify a pion as an electron 
was $\sim 0.8\%$, independent of pion momentum.

Muon ID employs two planes of
scintillation counters \cite{muon}.
The first plane (5.5 m $\times$ 3.0 m) of 15 counters 
measured the horizontal position while the second plane (3.0 m $\times$ 
2.2 m) of 16 counters measured the vertical position. There were 
about 15 interaction lengths of shielding upstream of the counters to 
filter out hadrons. 
Data from $D^+\rightarrow \overline{K}^{*0} \mu^{+}\nu _{\!_{\mu}}$ 
decays \cite{Chong} were used to choose selection criteria for muon 
candidates. 
Timing information from the smaller set of muon scintillation counters 
was used to improve the horizontal position resolution. Counter 
efficiencies, measured using muons originating from the primary target, 
were found to be $(99\pm 1)\%$ for the smaller counters and 
$(69\pm 3)\%$ for the larger counters. The probability of 
misidentifying a pion as a muon decreased with increasing momentum, 
from about 6$\%$ at 8 GeV/$c$ to $1.3\%$ above 20 GeV/$c$.
                                                                       
Events with evidence of well-separated production (primary) and decay 
(secondary) vertices were selected to separate charm candidates from 
background. Secondary vertices were required to be separated from the 
primary vertex by greater than $20\,\sigma_{_{\!L}}$ for $D^+$ decays 
and greater than $12\,\sigma_{_{\!L}}$ for $D^0$ and $D_{s}^{+}$ 
decays, where $\sigma_{_{\!L}}$ is the calculated resolution of the 
measured longitudinal separation. Also, the secondary vertex had to be 
separated from the closest material in the target foils by greater than 
$5\,\sigma_{_{\!L}}^{\prime }$, where $\sigma_{_{\!L}}^{\prime }$ is 
the uncertainty in this separation. The vector sum of the momenta 
from secondary vertex tracks was required to pass within 
$40~\mu$m of the primary vertex in the plane perpendicular to the beam. 
The net momentum of the charm candidate transverse to the line 
connecting the production and decay vertices had to be 
less than 300 MeV/$c$ for $D^0$ candidates,
less than 250 MeV/$c$ for $D_{s}^{+}$ candidates, and
less than 200 MeV/$c$ for $D^+$ candidates. 
Finally, decay track candidates were required to pass 
approximately 10 times closer to the secondary vertex than to the 
primary vertex. These selection criteria and kaon identification 
requirements were the same for both the search mode and for its 
normalization signal (discussed below).

To determine our selection criteria, we used a {\it blind} analysis 
technique. Before the selection criteria were finalized, all events 
having masses within a window $\Delta M_S$ around the mass of the 
$D^{0}$ were ``masked'' so that the presence or absence of any 
potential signal candidates would not bias our choice of selection 
criteria. All criteria were then chosen by studying events generated 
by a Monte Carlo (MC) simulation program \cite{MC} and background 
events, outside the signal windows, from real data. The criteria were 
chosen to maximize the ratio $N_{MC}/\sqrt{N_B}$, where $N_{MC}$ and 
$N_B$ are the numbers of MC and 
background events, respectively, after all selection criteria were 
applied. The data within the signal windows were unmasked only after 
this optimization. We used asymmetric windows for the decay modes 
containing electrons to allow for the bremsstrahlung low-energy tail. 


The upper limit for each branching fraction $B_{X}$ was calculated 
using the following formula: 
\begin{equation}
B_{X}=\frac{N_{X}}{N_{\mathrm{Norm}}}
\frac{\varepsilon _{\mathrm{Norm}}}{\varepsilon _{X}}
\times B_{\mathrm{Norm}};~\mathrm{where}~ 
\frac{\varepsilon _{\mathrm{Norm}}}{\varepsilon _{X}}=
\frac{f_{\mathrm{Norm}}^{\mathrm{MC}}}{f_{X}^{\mathrm{MC}}}.
\label{BReqn}
\end{equation}
$N_{X}$ is the 90$\%$ confidence level (CL) upper limit on the 
number of decays for the rare or forbidden decay mode $X$ 
and $B_{\mathrm{Norm}}$ is the normalization mode branching 
fraction obtained from the Particle Data Group \cite{PDG2000}. 
$\varepsilon_{\mathrm{Norm}}$ and $\varepsilon_{X}$ are the detection 
efficiencies while 
$f_{\mathrm{Norm}}^{\mathrm{MC}}$ and $f_{X}^{\mathrm{MC}}$ are 
the fractions of Monte Carlo events that were reconstructed and passed 
the final selection criteria, for the normalization and decay modes, 
respectively. 

The 90$\%$ CL upper limits $N_{X}$ are calculated using the method of 
Feldman and Cousins \cite{Feldman} to account for background, and then 
corrected for systematic errors by the method of Cousins and Highland 
\cite{Cousins}. In these methods, the numbers of signal events are 
determined by simple counting, not by a fit. Upper limits are 
determined using the number of candidate events observed and 
expected number of background events within the 
signal region.

Systematic errors include:
statistical errors from the fit to the normalization sample
$N_{\mathrm{Norm}}$; statistical errors on the numbers of Monte Carlo
events for both $N_{\mathrm{Norm}}^{\mathrm{MC}}$ and
$N_{X}^{\mathrm{MC}}$; uncertainties in the calculation of
mis-ID background; and uncertainties in the relative
efficiency for each mode, including lepton and kaon tagging.
These tagging efficiency uncertainties include: 1) the
muon counter efficiencies from both Monte Carlo simulation and
hardware performance; 2) kaon \cerenkov{} ID
efficiency due to differences in kinematics and modeling between
data and Monte Carlo simulated events; and 3) the fraction of
signal events (based on simulations) that would remain outside the
signal window due to bremsstrahlung tails.
The large systematic errors for the $D_{s}^{+}$ modes
are due to the uncertainty
in the branching fraction for the $D_{s}^{+}$ normalization mode.
The sums, taken in quadrature, of these systematic errors are listed
in Table \ref{Results1} and Table \ref{Results2}.

%
%
\begin{figure}[hb!]
\vskip -1.7 cm
\centerline{\epsfxsize 5.0 truein \epsfbox{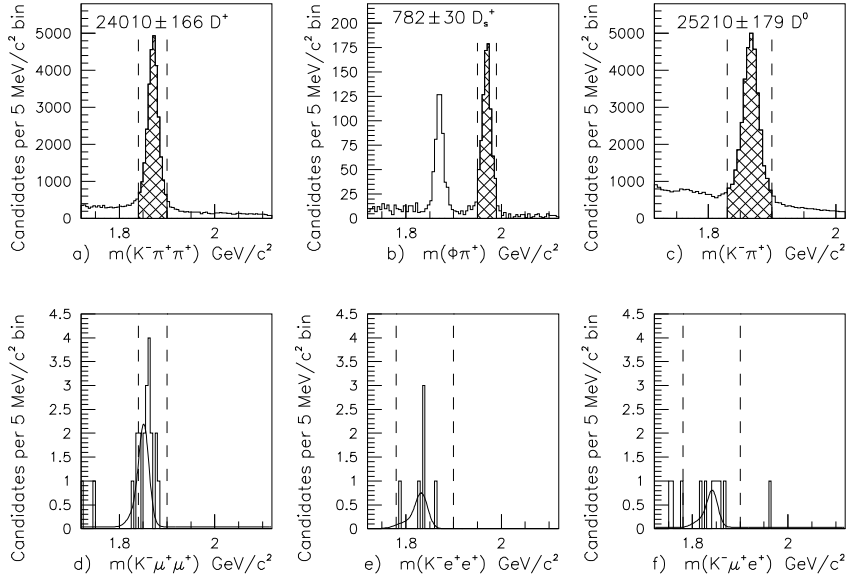}}
\vskip -.2 cm
\vspace*{-11pt}
\caption[]{
\small
Top row: typical normalization charm signals.
The signal region is
shaded.
Bottom row: invariant mass plots of $D^{+}$ candidate decays to
$K^-\mu ^+\mu ^+$, $K^-e^+e^+$, and $K^-\mu ^+e^+$, showing
reflections mostly from misidentified $D^+\rightarrow
K^-\pi^+\pi^+$ decays. These modes are used to set mis-ID rate
rather than upper limits
The solid curves are normalized
Monte Carlo fits. The dashed lines show the signal window.
}
\label{Normal}
\end{figure}
\vskip 20pt 
\leftline{\bf The $D^+\rightarrow h\ell \ell $, 
$D_{s}^+\rightarrow h\ell \ell $, and $D^0\rightarrow \ell^+\ell^-$ 
Analysis}
\vskip 3pt
We normalized the sensitivity of our search to the topologically similar 
Cabibbo-favored decays shown in Table \ref{Norm1} and Figure \ref{Normal}. 
For the $D^{+}$ decays we used 
$D^+\rightarrow K^-\pi^+\pi^+$; for $D_{s}^{+}$ 
decays we used $D_{s}^{+}\rightarrow \phi \pi^+$; and for 
$D^{0}$ decays we used $D^0\rightarrow K^-\pi^+$ events. 
The efficiencies for the normalization modes varied from about $0.5\%$ to
$2\%$.

\begin{table}[th!]
\vspace*{-5pt}
\caption[]{
Normalization modes used for 
$D^+\rightarrow h\ell \ell $, 
$D_{s}^+\rightarrow h\ell \ell $, and $D^0\rightarrow \ell^+\ell^-$.}
\label{Norm1}
\tabcolsep=1.5mm
\begin{center}
\renewcommand{\arraystretch}{1.05}
\begin{tabular}{lllcl}
\hline
\vspace*{-10pt} &  &  & &  \\
Rare $D$ Decay & $D$ Norm. & Events & MC Efficiency 
& PDG98 \cite{PDG98} Branching Ratio \\
\hline
\vspace*{-10pt} &  &  & &  \\
$D^+\to \pi\ell \ell $ & $D^+\to K^- \pi^+ \pi^+$ & 24010$\pm$166 & 
 1.06\% & $(9.0 \pm 0.6) \%$ \\
$D^+\to K\ell \ell $ & $D^+\to K^- \pi^+ \pi^+$ & 17730$\pm$141 & 
 0.82\% & $(9.0 \pm 0.6) \%$ \\
$D_{s}^+\to \pi\ell \ell $   & $D_{s}^+\to \phi \pi^+  $ & 952$\pm$34 &
 0.60\% & (3.6 $\pm$ 0.9)\% \\
$D_{s}^+\to K^+\ell^+ \ell^- $   & $D_{s}^+\to \phi \pi^+  $ & 782$\pm$30 &
 0.46\% & (3.6 $\pm$ 0.9)\% \\
$D_{s}^+\to K^-\ell^+ \ell^+ $   & $D_{s}^+\to \phi \pi^+  $ & 679$\pm$27 &
 0.49\% & (3.6 $\pm$ 0.9)\% \\
$D^0\rightarrow \ell^+\ell^-$  & $D^0\rightarrow K^- \pi^+$ & 25210$\pm$179 &
 1.81\% & (3.85 $\pm$ 0.09)\% \\
\hline
\end{tabular}
\end{center}
\vspace*{-5pt}
\end{table}

The widths of our normalization modes were 10.5 MeV/$c^{\,2}$ for 
$D^{+}$, 9.5 MeV/$c^{\,2}$ for $D_{s}^{+}$, and 12 MeV/$c^{\,2}$ for 
$D^{0}$. 
The signal windows used are: 

\begin{table}[ht!]
\begin{center}
\renewcommand{\arraystretch}{1.05}
\vspace*{-6pt}
\begin{tabular}{ll}
$1.84<M(D^{+})<1.90 ~{\rm for}~ D^{+}\rightarrow h\mu \mu$ \hfil &
$1.78<M(D^{+})<1.90 ~{\rm GeV}/c^{\,2}~{\rm for}~ D^{+}\rightarrow hee ~{\rm
and}~  h\mu e$ \\
$1.95<M(D_{s}^{+})<1.99 ~{\rm for}~ D_{s}^{+}\rightarrow
h\mu \mu$ \hfil &
$1.91<M(D_{s}^{+})<1.99 ~{\rm GeV}/c^{\,2}~{\rm for}~
D_{s}^{+}\rightarrow hee~{\rm and}~ h\mu e$ \\
$1.83<M(D^{0})<1.90 ~{\rm for}~
D^{0}\rightarrow \mu \mu$ \hfil &
$1.76<M(D^{0})<1.90 ~{\rm GeV}/c^{\,2}~{\rm for}~
D^{0}\rightarrow ee~{\rm and}~ \mu e$ \\
\end{tabular}
\vspace*{-5pt}
\end{center}
\end{table}
\begin{figure}[t!]
\vskip -2.5 cm
\centerline{\epsfxsize 5.0 truein \epsfbox{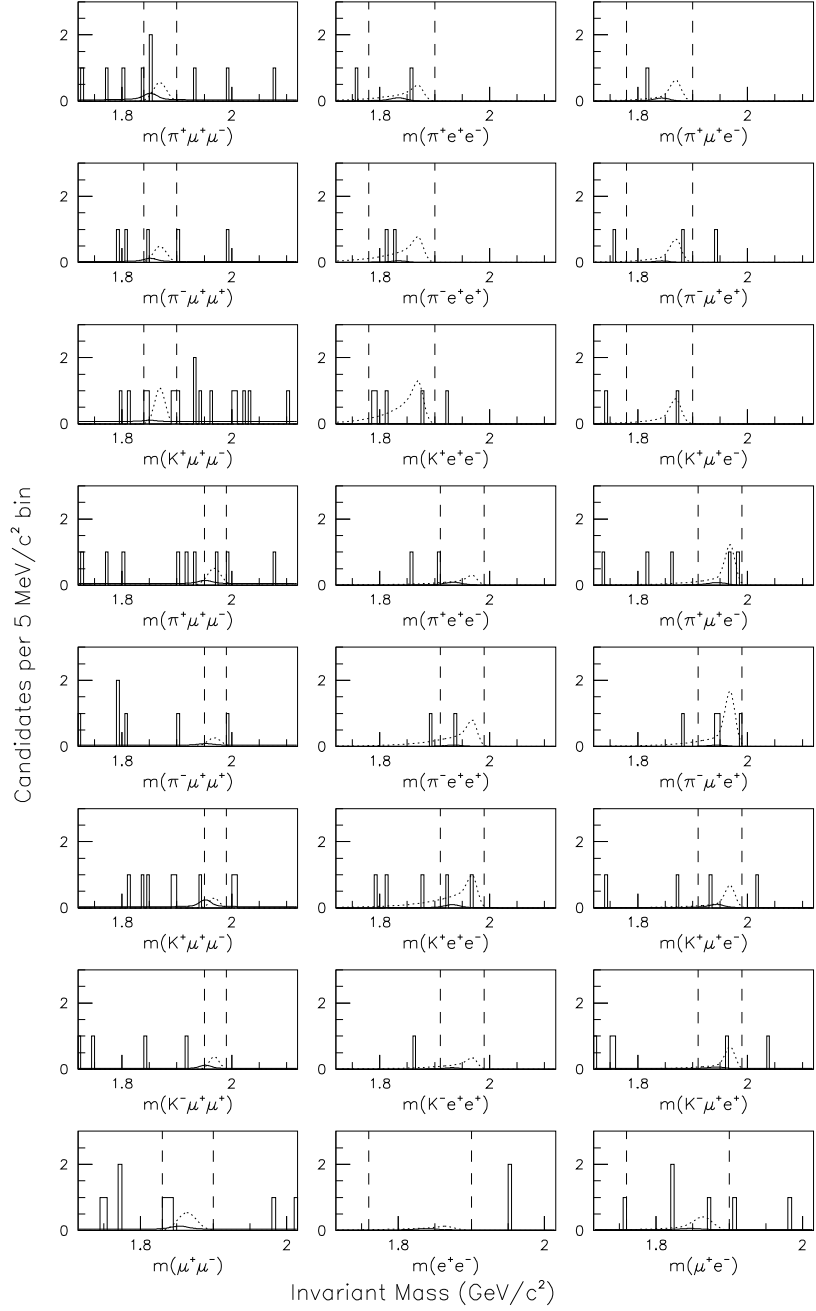}}
\vskip -.2 cm
\vspace*{-11pt}
\caption[]{
\small Final event samples for the $D^+$ (rows 1--3),
$D_{s}^{+}$ (rows 4--7), and $D^0$ (row 8) decays. The solid curves
represent estimated background; the dotted curves represent signal
shape for a number of events equal to the 90$\%$ CL upper limit.
The dashed vertical lines are $\Delta M_S$ boundaries.}
\label{Data1}
\vspace*{-12pt}
\end{figure}
Background that is not removed by cuts
include decays in which hadrons (from
real, fully-hadronic decay vertices) are misidentified as leptons.
In the case where kaons are misidentified as leptons, candidates
have effective masses which lie outside the signal windows. Most of
these originate from Cabibbo-favored modes
$D^+\rightarrow K^-\pi^+\pi^+$, $D_{s}^{+}\rightarrow K^-K^+\pi^+$,
and $D^0\rightarrow K^-\pi^+$. These
Cabibbo-favored reflections were explicitly removed prior to
cut optimization. There remain two sources of background
in our data: hadronic decays with pions misidentified as leptons
($N_{\mathrm{MisID}}$) and ``combinatoric'' background
($N_{\mathrm{Cmb}}$) arising primarily from false vertices and
partially reconstructed charm decays. After cuts
were applied and the signal windows opened, the number of events
within the window is $N_{\mathrm{Obs}} = N_{\mathrm{Sig}}+
N_{\mathrm{MisID}} + N_{\mathrm{Cmb}}$.

%
\begin{table}[th!]
\caption[]{
E791 90$\%$ confidence level (CL) branching fractions (BF) compared
to previous experiments.
The background and candidate events
correspond to the signal region only.
The Monte Carlo (MC)
yield is from 250\,000 generated events in each of the 24 cases.
}
\label{Results1}
\vskip 5pt
\tabcolsep=4.0pt
\begin{center}
\begin{tabular}{lcccccrlll} \hline
&(Est.&BG)&Cand.&Syst.&90$\%$ CL & MC &E791&Previous & Experi-  \\
Mode&$N_{\mathrm{Cmb}}$&$N_{\mathrm{MisID}}$&Obs.&Err.&Num. & Yield
&$BF$ Limit&$BF$ Limit & ment \\
\hline
\vspace*{-11pt} &     &   &     &     &         &    &         &     \\
 $D^{+}\rightarrow \pi ^{+}\mu ^{+}\mu ^{-}$&1.20&1.47&2&10$\%$&3.35    &2706
 &$1.5\times 10^{-5}$&$1.8\times 10^{-5}$ & E791 \cite{FCNC} \\
 $D^{+}\rightarrow \pi ^{+}e^{+}e^{-}$&0.00&0.90&1&12$\%$&3.53          &816
 &$5.2\times 10^{-5}$&$6.6\times 10^{-5}$ & E791 \cite{FCNC} \\
 $D^{+}\rightarrow \pi ^{+}\mu ^{\pm }e^{\mp }$&0.00&0.78&1&11$\%$&3.64 &1272
 &$3.4\times 10^{-5}$&$1.2\times 10^{-4}$ & E687 \cite{E687} \\
 $D^{+}\rightarrow \pi ^{-}\mu ^{+}\mu ^{+}$&0.80&0.73&1&9$\%$&2.92     &2088
 &$1.7\times 10^{-5}$&$8.7\times 10^{-5}$ & E687 \cite{E687} \\
 $D^{+}\rightarrow \pi ^{-}e^{+}e^{+}$&0.00&0.45&2&12$\%$&5.60          &701
 &$9.6\times 10^{-5}$&$1.1\times 10^{-4}$ & E687 \cite{E687} \\
 $D^{+}\rightarrow \pi ^{-}\mu ^{+}e^{+}$&0.00&0.39&1&11$\%$&4.05       &976
 &$5.0\times 10^{-5}$&$1.1\times 10^{-4}$ & E687 \cite{E687} \\
 $D^{+}\rightarrow K^{+}\mu ^{+}\mu ^{-}$&2.20&0.20&3&8$\%$&5.07        &1206
 &$4.4\times 10^{-5}$&$9.7\times 10^{-5}$ & E687 \cite{E687} \\
 $D^{+}\rightarrow K^{+}e^{+}e^{-}$&0.00&0.09&4&11$\%$&8.72             &453
 &$2.0\times 10^{-4}$&$2.0\times 10^{-4}$ & E687 \cite{E687} \\
 $D^{+}\rightarrow K^{+}\mu ^{\pm }e^{\mp }$&0.00&0.08&1&9$\%$&4.34     &664
 &$6.8\times 10^{-5}$&$1.3\times 10^{-4}$ & E687 \cite{E687} \\
\hline
\vspace*{-11pt} &     &   &     &     &         &    &         &     & \\
 $D_{s}^{+}\rightarrow K^{+}\mu ^{+}\mu ^{-}$&0.67&1.33&0&27$\%$&1.32   &647
 &$1.4\times 10^{-4}$&$5.9\times 10^{-4}$ & E653 \cite{E653} \\
 $D_{s}^{+}\rightarrow K^{+}e^{+}e^{-}$&0.00&0.85&2&29$\%$&5.77         &244
 &$1.6\times 10^{-3}$& & \\
 $D_{s}^{+}\rightarrow K^{+}\mu ^{\pm }e^{\mp }$&0.40&0.70&1&27$\%$&3.57&388
 &$6.3\times 10^{-4}$& & \\
 $D_{s}^{+}\rightarrow K^{-}\mu ^{+}\mu ^{+}$&0.40&0.64&0&26$\%$&1.68   &686
 &$1.8\times 10^{-4}$&$5.9\times 10^{-4}$  & E653 \cite{E653} \\
 $D_{s}^{+}\rightarrow K^{-}e^{+}e^{+}$&0.00&0.39&0&28$\%$&2.22         &257
 &$6.3\times 10^{-4}$& & \\
 $D_{s}^{+}\rightarrow K^{-}\mu ^{+}e^{+}$&0.80&0.35&1&27$\%$&3.53      &381
 &$6.8\times 10^{-4}$& & \\
 $D_{s}^{+}\rightarrow \pi ^{+}\mu ^{+}\mu ^{-}$&0.93&0.72&1&27$\%$&3.02&1725
 &$1.4\times 10^{-4}$&$4.3\times 10^{-4}$  & E653 \cite{E653} \\
 $D_{s}^{+}\rightarrow \pi ^{+}e^{+}e^{-}$&0.00&0.83&0&29$\%$&1.85      &565
 &$2.7\times 10^{-4}$& & \\
 $D_{s}^{+}\rightarrow \pi^+ \mu^{\pm}e^{\mp}$&0.00&0.72&2&30$\%$&6.01  &809
 &$6.1\times 10^{-4}$& & \\
 $D_{s}^{+}\rightarrow \pi ^{-}\mu ^{+}\mu ^{+}$&0.80&0.36&0&27$\%$&1.60&1588
 &$8.2\times 10^{-5}$&$4.3\times 10^{-4}$  & E653 \cite{E653} \\
 $D_{s}^{+}\rightarrow \pi ^{-}e^{+}e^{+}$&0.00&0.42&1&29$\%$&4.44      &528
 &$6.9\times 10^{-4}$& &\\
 $D_{s}^{+}\rightarrow \pi ^{-}\mu ^{+}e^{+}$&0.00&0.36&3&28$\%$&8.21   &911
 &$7.3\times 10^{-4}$& & \\
\vspace*{-12pt} &     &   &     &     &         &    &         &     &\\
\hline
\vspace*{-11pt} &     &   &     &     &         &    &         &     &\\
 $D^{0}\rightarrow \mu ^{+}\mu ^{-}$&1.83&0.63&2&6$\%$&3.51             &5297
 &$5.2\times 10^{-6}$&$4.1\times 10^{-6}$  & WA92 \cite{E771} \\
 $D^{0}\rightarrow e^{+}e^{-}$&1.75&0.29&0&9$\%$&1.26                   &1577
 &$6.2\times 10^{-6}$&$8.2\times 10^{-6}$ & E789 \cite{E789} \\
 $D^{0}\rightarrow \mu ^{\pm }e^{\mp }$&2.63&0.25&2&7$\%$&3.09          &2983
 &$8.1\times 10^{-6}$&$1.7\times 10^{-5}$  & E789 \cite{E789} \\
\hline
\end{tabular}
\end{center}
\end{table}
%

The background $N_{\mathrm{MisID}}$ arises mainly
from singly-Cabibbo-suppressed (SCS) modes. These misidentified
leptons can come from hadronic shower punchthrough,
decays-in-flight, and random overlaps of
tracks. We do
not attempt to establish a limit for $D^+\rightarrow K^-\ell^+\ell^+$
modes, as they have relatively large feedthrough signals from copious
Cabibbo-favored $K^-\pi^+\pi^+$ decays. Instead, we use the observed
signals in $K^-\ell^+\ell^+$ channels to measure three dilepton
mis-ID rates under the assumption that the observed signals
(shown in Figs.~\ref{Normal}d--f) arise entirely from lepton
mis-ID. The curve shapes are from Monte Carlo.
The following mis-ID rates were obtained:
$r_{\mu\mu}= (7.3 \pm 2.0)\times 10^{-4}$,
$r_{\mu e}= (2.9 \pm 1.3 )\times 10^{-4}$, and
$r_{e e}= (3.4 \pm 1.4)\times 10^{-4}$.
Using these rates we estimate the numbers of misidentified candidates,
$N_{\mathrm{MisID}}^{h\ell\ell}$ (for $D^{+}$ and $D_{s}^{+}$) and
$N_{\mathrm{MisID}}^{\ell\ell}$ (for $D^{0}$), in the signal windows
as follows:
$N_{\mathrm{MisID}}^{h\ell\ell} = r_{\ell\ell}
\cdot N_{\mathrm{SCS}}^{h\pi\pi}$
and
$N_{\mathrm{MisID}}^{\ell\ell} = r_{\ell\ell}
\cdot N_{\mathrm{SCS}}^{\pi\pi}$,
where $N_{\mathrm{SCS}}^{h\pi\pi}$ and $N_{\mathrm{SCS}}^{\pi\pi}$
are the numbers of SCS hadronic decay candidates within the signal
windows. For modes in which two possible pion combinations can
contribute, \eg, $D^+\rightarrow h^{+}\mu ^{\pm}\mu ^{\mp}$, we double
the rate.

To estimate the combinatoric background $N_{\mathrm{Cmb}}$ within a
signal window $\Delta M_S$, we count events having masses within an
adjacent background mass window $\Delta M_B$, and scale this number
($N_{\Delta M_B}$) by the relative sizes of these windows:
$N_{\mathrm{Cmb}} = ({\Delta M_S}/{\Delta M_B}) \cdot N_{\Delta M_B}$.
To be conservative in calculating our 90$\%$ confidence level upper
limits, we take combinatoric backgrounds to be zero when no
events are located above the mass windows.

\begin{figure}[t!]
   \begin{minipage}{3.15in}
    \centerline{\epsfxsize 3.15 truein \epsfbox{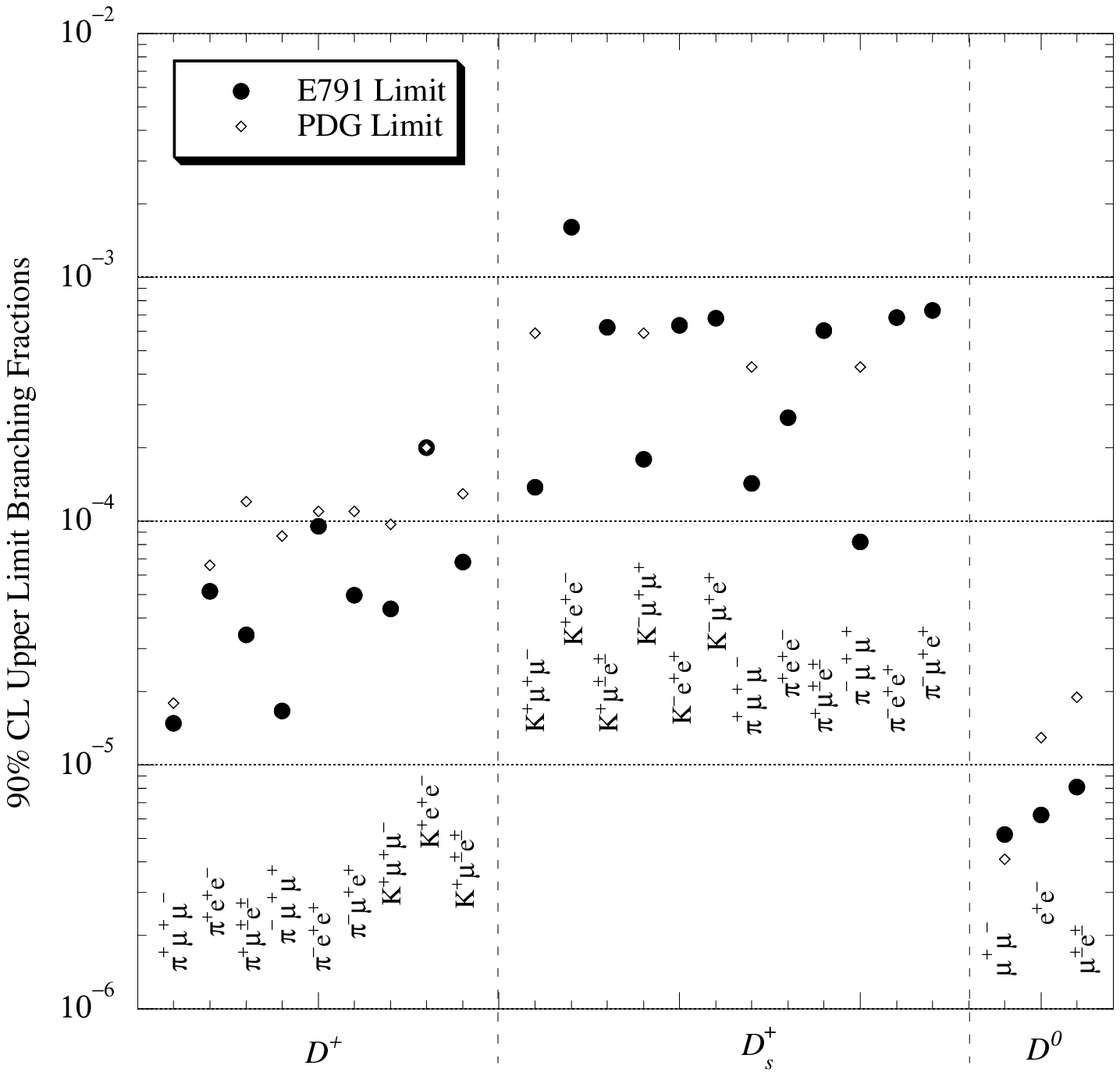}}
    \caption{
\small 
Comparison of the 90$\%$ CL upper-limit branching fractions 
from E791 data (dark circles) with existing limits (open diamonds) from 
the 1998 PDG [44].  The modes are $D^+$ or $D_s^+ \to \pi \ell \ell$
or $K \ell \ell$ and $D^0 \to \ell^+ \ell^-$.
}
    \label{BR1}
   \end{minipage}
\hspace{0.4in}
   \begin{minipage}{3.05in}
    \centerline{\epsfxsize 3.05 truein \epsfbox{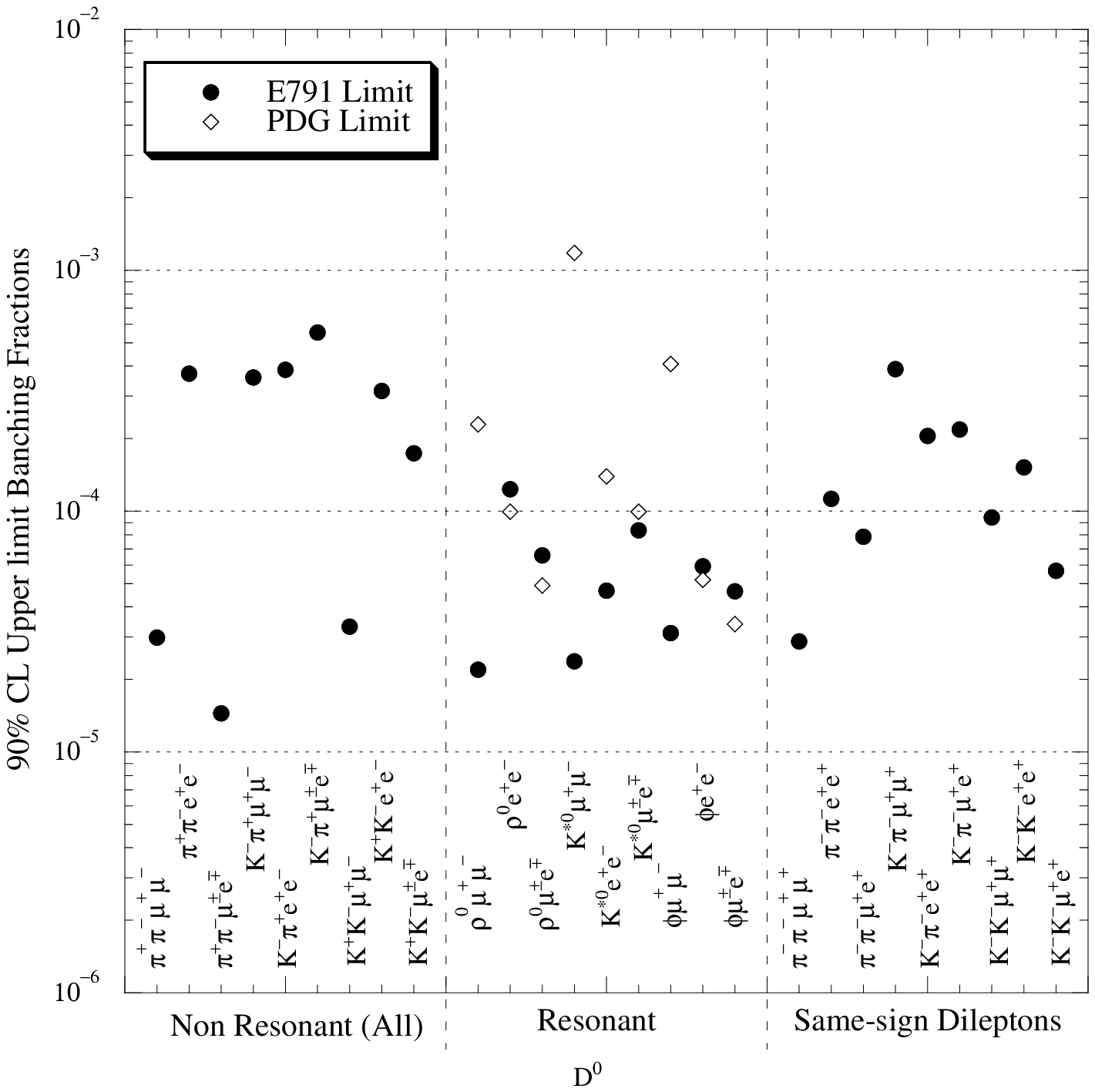}}
    \caption{
\small Comparison of the 90$\%$ CL upper-limit 
$D^0 \to V\ell^+\ell^-$ and $h h \ell \ell$
branching fractions 
from E791 data (dark circles) with existing limits (open diamonds) from 
the 2000 PDG [51]. 
}
    \label{BR2}
   \end{minipage}
\end{figure}

In Table \ref{Results1} we
present the numbers of combinatoric background, mis-ID
background, and observed events for all 24 modes.
The efficiencies for the search modes varied from about $0.1\%$ to $2\%$.
Data are shown in Figure \ref{Data1} and limits are compared 
with previous results in Figure \ref{BR1}.

\medskip
\leftline{\bf The $D^0\rightarrow V\ell^+\ell^-$ and 
$D^0\rightarrow hh\ell \ell$ Analysis}
\vskip 3pt
There were a few minor differences between this analysis and our 
previous analysis as discussed above. First, we examined resonant 
modes, where the mass ranges used were: 
$\left| m_{\pi ^+\pi ^-} - m_{\rho ^0}\right| <150$ MeV/$c^2$, 
$\left| m_{K^-\pi ^+} - m_{\Kstar}\right| <55$ MeV/$c^2$, and 
$\left| m_{K^+K^-} - m_{\phi}\right| <10$ MeV/$c^2$.
We normalized the sensitivity of each search to 
similar hadronic 3-body (resonant) or 4-body (non-resonant) decays. 
One exception is the case of 
$D^{0}\rightarrow \rho^{0} \ell ^{\pm }\ell ^{\mp }$ where we 
normalize to nonresonant  
$D^{0}\rightarrow \pi ^+\pi ^-\pi ^+\pi ^-$ because no published 
branching fraction exists for 
$D^0\rightarrow \rho ^{0}\pi ^+\pi ^-$. Table \ref{Norm2} lists the 
normalization mode used for each signal mode and 
the fitted numbers of normalization data events ($N_{\mathrm{Norm}}$).
The efficiencies for the normalization modes varied from
$0.2\%$ to $1\%$, and the efficiencies for the
search modes varied from $0.05\%$ to $0.34\%$.
The signal windows are:
$1.83<M(D^0)<1.90$ GeV/$c^2$ for $\mu \mu $ and
$1.76<M(D^0)<1.90$ GeV/$c^2$ for $ee$ and $\mu e$ modes.

\begin{table}[th!]
\vspace*{-5pt}
\caption[]{
Normalization modes used for $D^0\rightarrow V\ell^+\ell^-$ and 
$D^0\rightarrow hh\ell \ell$.}
\label{Norm2}
\tabcolsep=1.5mm
\begin{center}
\begin{tabular}{lllcl}
\hline
\vspace*{-10pt} &  &  & &  \\
Rare $D^0$ Decay & $D^0$ Norm. & Events & MC Efficiency 
& PDG2000 \cite{PDG2000} Branching Ratio \\
\hline
\vspace*{-10pt} &  &  & &  \\
$\rho^{0} \ell ^{\pm }\ell ^{\mp }$& $\pi ^+\pi ^-\pi ^+\pi ^-$& 2049$\pm$53 & 
 0.95\% & $(7.3 \pm 0.5) \times 10^{-3}$ \\
$\Kstar \ell ^{\pm }\ell ^{\mp }$& $\Kstar \pi ^+\pi ^-$& 5451$\pm$72 &
 0.28\% & (1.4 $\pm$ 0.4)\% \\
$\phi \ell ^{\pm }\ell ^{\mp }$& $\phi \pi ^+\pi ^-$& 113$\pm$19 &
 0.21\% & $(1.07 \pm 0.28) \times 10^{-3}$\\
$\pi \pi \ell\ell $& $\pi ^+\pi ^-\pi ^+\pi ^-$& 2049$\pm$53 &
 0.95\% & $(7.3 \pm 0.5) \times 10^{-3}$ \\
$K\pi \ell\ell $& $K^-\pi ^+\pi ^-\pi ^+$&11550$\pm$113 &
 0.41\% & (7.49 $\pm$ 0.31)\% \\
$KK\ell\ell $& $K^+K^-\pi ^+\pi ^-$& 406$\pm$41 &
 0.26\% & $(2.5 \pm 0.23) \times 10^{-3}$ \\
\hline
\end{tabular}
\end{center}
\vspace*{-5pt}
\end{table}

Background sources that are not removed by the selection criteria
discussed earlier include decays in which hadrons (from real,
fully-hadronic decay vertices) are misidentified as leptons. These
misidentified leptons can come from hadronic showers reaching muon
counters, decays-in-flight, and random overlaps of tracks from
otherwise separate decays (``accidental'' sources). In the case where
kaons are misidentified as pions or leptons, candidate masses shift
below signal windows. However, we remove these events to prevent
them from influencing our background estimate, which is
partially obtained from the mass sidebands (see discussion of
$N_{\mathrm{Cmb}}$ below). To remove these events prior to the
selection-criteria optimization, we reconstruct all candidates as each
of the non-resonant normalization modes and test whether the
masses are consistent with $m_{D^{0}}$. If so, we remove the events,
but only if the number of kaons in the final state differs from that
of the search mode. We do not remove events having the same number of
kaons, as the loss in acceptance for true signal events would be
excessive.

\begin{figure}[b!]
\vspace*{-18pt}
\centerline{\epsfxsize 3.375 truein \epsfbox{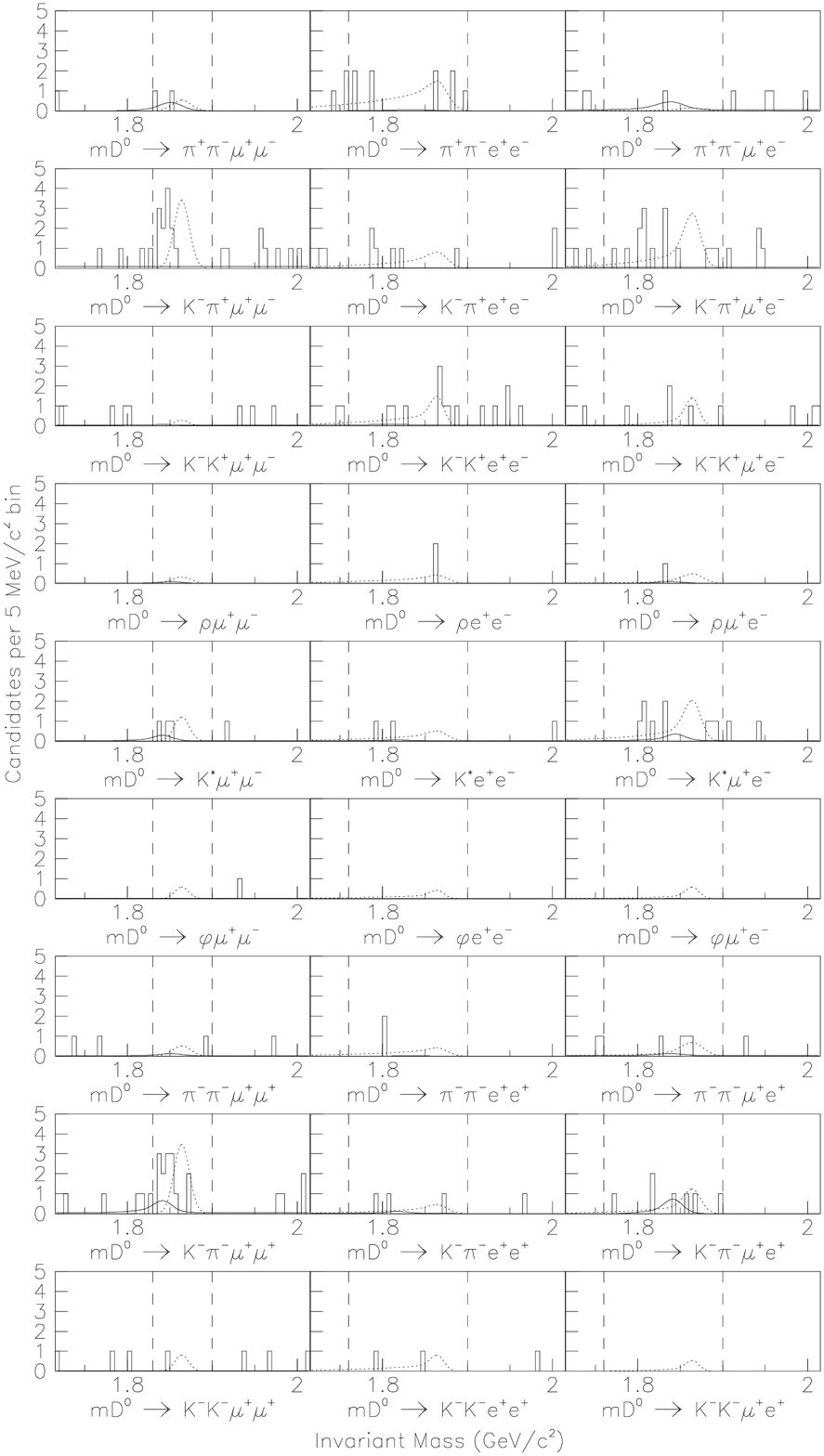}}
\caption[]{
\small Final event samples for the opposite signed dilepton (rows 1--3),
resonant (rows 4--6), and same signed dilepton modes (rows 7--9) of
$D^0$ decays. The solid curves display total estimated background; the
dotted curves display signal shape for a number of events equal to
the 90$\%$ CL upper limit. The dashed vertical lines are the
$\Delta M_S$ boundaries.}
\label{Data_4prong}
\end{figure}

There remain two sources of background: hadronic decays where pions are
misidentified as leptons ($N_{\mathrm{MisID}}$) and ``combinatoric''
background ($N_{\mathrm{Cmb}}$) arising primarily from false vertices
and partially reconstructed charm decays. The background
$N_{\mathrm{MisID}}$ arises from the normalization modes. To estimate
the rate for misidentifying $\pi \pi $ as $\ell \ell $, for all but the
$D^0\rightarrow K^-\pi ^+\ell ^+\ell ^-$ modes, we assume all
$D^0\rightarrow K^-\pi ^+\ell ^+\ell ^-$ candidates observed (after
subtracting combinatoric background estimated from mass sidebands)
result from misidentification of $D^0\rightarrow K^-\pi ^+\pi ^-\pi ^+$
decays and count the number of $D^0\rightarrow K^-\pi ^+\ell ^+\ell ^-$
decays passing the final selection criteria. We then divide by twice
the number of $D^0\rightarrow K^-\pi ^+\pi ^-\pi ^+$ normalization
events with $K^-\pi ^+\ell ^+\ell ^-$ mass within $\Delta M_S$
boundaries (twice because there are two possible $\pi ^+$
misidentifications).

%
%
\begin{table}[b!]
\begin{center}
\vspace*{-6pt}
\caption[]{
E791 90$\%$ confidence level (CL) upper limits on the number of events
and branching fraction limits ($\!\times \!10^{-5}$). The Monte Carlo (MC)
yield is from 250\,000 generated events in each of the 27 cases. Previously
published limits are listed for comparison \cite{PDG2000,CLEO,E653}.
}
\label{Results2}
\vskip 1pt
\tabcolsep=2.0pt
\begin{tabular}{lcccccrrr}
&(Est.&BG)&&Sys.&&MC & \small{E791}&\small{PDG}\\
Mode $D^{0}\rightarrow $&\small{$N_{\mathrm{Cmb}}$}&
\small{$N_{\mathrm{MisID}}$}&\small{$N_{\mathrm{Obs}}$}&
Err.&\small{$N_{X}$}&Yield&\small{Limit}&\small{Limit}\\
\hline
\vspace*{-10pt} &    &   &     &     &    &    &   &\\
 $\pi ^{+}\pi ^{-}\mu ^{+}\mu ^{-}$&0.00&3.16&2&11$\%$&2.96   &840 & $3.0$&\\
 $\pi ^{+}\pi ^{-}e^{+}e^{-}$&0.00&0.73&9&12$\%$&15.2         &345 &$37.3$&\\
 $\pi ^{+}\pi ^{-}\mu ^{\pm }e^{\mp }$&5.25&3.46&1&15$\%$&1.06&620 & $1.5$&\\
 $K^{-}\pi ^{+}\mu ^{+}\mu ^{-}$&3.65&0.00&12&11$\%$&15.4     &286 &$35.9$&\\
 $K^{-}\pi ^{+}e^{+}e^{-}$&3.50&0.00&6&15$\%$&7.53            &135 &$38.5$&\\
 $K^{-}\pi ^{+}\mu ^{\pm }e^{\mp }$&5.25&0.00&15&12$\%$&17.3  &217 &$55.3$&\\
 $K^{+}K^{-}\mu ^{+}\mu ^{-}$&2.13&0.17&0&17$\%$&1.22         &145 & $3.3$&\\
 $K^{+}K^{-}e^{+}e^{-}$&6.13&0.04&9&18$\%$&9.61               &120 &$31.5$&\\
 $K^{+}K^{-}\mu ^{\pm }e^{\mp }$&3.50&0.17&5&17$\%$&6.61      &149 &$18$  &\\
\hline
\vspace*{-10pt} &    &   &     &     &    &    &   &\\
 $\rho ^{0}\mu ^{+}\mu ^{-}$&0.00&0.75&0&10$\%$&1.80      &694 & $2.2$&$23$ \\
 $\rho ^{0}e^{+}e^{-}$&0.00&0.18&1&12$\%$&4.28            &294 &$12.4$&$10$ \\
 $\rho ^{0}\mu ^{\pm }e^{\mp }$&0.00&0.82&1&11$\%$&3.60   &466 &$6.6$ &$4.9$\\
 $\Kstar \mu ^{+}\mu ^{-}$&0.30&1.87&3&24$\%$&5.40        &275 &$2.4$ &$118$\\
 $\Kstar e^{+}e^{-}$&0.88&0.49&2&25$\%$&4.68              &121 &$4.7$ &$14$ \\
 $\Kstar \mu ^{\pm }e^{\mp }$&1.75&2.30&9&24$\%$&12.8     &185 &$8.3$ &$10$ \\
 $\phi \mu ^{+}\mu ^{-}$&0.30&0.04&0&33$\%$&2.33          &187 &$3.1$ &$41$ \\
 $\phi e^{+}e^{-}$&0.00&0.01&0&33$\%$&2.75                &117 &$5.9$ &$5.2$\\
 $\phi \mu ^{\pm }e^{\mp }$&0.00&0.05&0&33$\%$&2.71       &146 &$4.7$ &$3.4$\\
\hline
\vspace*{-10pt} &    &   &     &     &    &    &  & \\
 $\pi ^{-}\pi ^{-}\mu ^{+}\mu ^{+}$&0.91&0.79&1&9$\%$&2.78    &821   &$2.9$ &\\
 $\pi ^{-}\pi ^{-}e^{+}e^{+}$&0.00&0.18&1&11$\%$&4.26         &322   &$11.2$&\\
 $\pi ^{-}\pi ^{-}\mu ^{+}e^{+}$&2.63&0.86&4&10$\%$&5.18      &559   &$7.9$ &\\
 $K^{-}\pi ^{-}\mu ^{+}\mu ^{+}$&2.74&3.96&14&9$\%$&15.7      &268   &$39.0$&\\
 $K^{-}\pi ^{-}e^{+}e^{+}$&0.88&1.04&2&16$\%$&4.14            &134   &$20.6$&\\
 $K^{-}\pi ^{-}\mu ^{+}e^{+}$&0.00&4.88&7&11$\%$&7.81         &238   &$21.8$&\\
 $K^{-}K^{-}\mu ^{+}\mu ^{+}$&1.22&0.00&1&17$\%$&3.27         &137   &$9.4$ &\\
 $K^{-}K^{-}e^{+}e^{+}$&0.88&0.00&2&17$\%$&5.28               &137   &$15.2$&\\
 $K^{-}K^{-}\mu ^{+}e^{+}$&0.00&0.00&0&17$\%$&2.52            &175   &$5.7$ &\\
\end{tabular}
\end{center}
\end{table}
Table \ref{Results2} shows
numbers of combinatoric background, misidentification
background, and observed events for all 27 modes.
Data are shown in Figure \ref{Data_4prong} 
and compared with previous results in Figure \ref{BR2}.
%


From this procedure, the following misidentification rates were
obtained:
$r_{\mu\mu}= (3.4 \pm 2.4)\times 10^{-4}$,
$r_{\mu e}= (4.2 \pm 1.4)\times 10^{-4}$, and
$r_{ee}= (9.0 \pm 6.2 )\times 10^{-5}$.
For modes in which two possible pion combinations can contribute, \eg,
$D^0\rightarrow K^-\pi ^{+}\mu ^{\pm}\mu ^{\mp}$, we use twice the
above rate; and for $D^{0}\rightarrow \pi ^+\pi ^-\pi ^+\pi ^-$, where
there are 4 possible combinations, we use 4 times this rate in
calculating $D^{0}\rightarrow \pi ^+\pi ^-\ell ^+\ell ^-$.
Using these rates, we estimate the numbers of misidentified candidates,
$N_{\mathrm{MisID}}^{V\ell\ell}$ and $N_{\mathrm{MisID}}^{hh\ell\ell}$,
in the signal windows as follows:
\vspace*{-3pt}
\begin{equation}
N_{\mathrm{MisID}}^{hh\ell\ell} = r_{\ell\ell}
\times N_{\mathrm{Norm}}^{hh\pi\pi}
~{\rm and}~ N_{\mathrm{MisID}}^{V\ell\ell} = r_{\ell\ell}
\times N_{\mathrm{Norm}}^{V\pi\pi}~,
\label{Nmidid}
\end{equation}
where $N_{\mathrm{Norm}}^{hh\pi\pi}$ and $N_{\mathrm{Norm}}^{V\pi\pi}$
are the numbers of normalization hadronic decay candidates in the
signal windows.

To calculate the upper limits for the
$D^0\rightarrow K^-\pi ^+\ell ^+\ell ^-$ modes, we set
$N_{\mathrm{MisID}}$ to zero as we do not have an independent estimate
of the misidentification rates. This results in conservative upper
limits. If we had used the misidentification rates from our previous,
3-body decay study \cite{FCNCnew}, then our limits for the three
$D^0\rightarrow K^-\pi ^+\ell ^+\ell ^-$ modes would be lower by about
a factor of two.

To estimate the combinatoric background $N_{\mathrm{Cmb}}$ within a
signal window $\Delta M_S$, we count events having masses within an
adjacent background mass window $\Delta M_B$, and scale this number
($N_{\Delta M_B}$) by the relative sizes of these windows:
$N_{\mathrm{Cmb}} = ({\Delta M_S}/{\Delta M_B}) \times N_{\Delta M_B}$.
To be conservative in calculating our 90$\%$ confidence level upper
limits, we take combinatoric backgrounds to be zero when no
events are located above the mass windows.

\vskip 20pt 
\leftline{\bf Conclusion}
\vskip 3pt
We used a {\it blind} analysis of data from Fermilab 
experiment E791 to obtain upper limits on the dilepton branching 
fractions for 51 flavor-changing neutral current, lepton-number 
violating, and lepton-family violating decays of $D^+$, $D_{s}^{+}$, 
and $D^0$ mesons. No evidence for any of these 2, 3 and 4-body decays 
was found. Therefore, we presented upper limits on the branching 
fractions at the 90$\%$ confidence level. Eighteen limits represented 
significant improvements over previously published results. Twenty-six 
of the remaining modes had no previously reported limits.
Work is currently underway at the Fermilab FOCUS \cite{Sheldon} experiment and
others to further improve the limits presented here or to observe signals. 
%
\vskip 15pt
\leftline{\bf Acknowledgments}
\vskip 3pt

Many thanks to Eric Aitala \cite{aitala}, Jeff Appel \cite{appel}, 
Steve Bracker \cite{bracker}, Lucien Cremaldi \cite{cremaldi},
Ai Nguyen\cite{nguyen}, 
Milind Purohit \cite{purohit}, David Sanders \cite{sanders}, 
Alan Schwartz \cite{SCHWARTZ93}, 
Jean Slaughter \cite{slaughter},
Noel Stanton \cite{stanton}, 
Nick Witchey \cite{witchey},
and all 
the members of the Fermilab E791 collaboration.
Also many thanks to the staffs of Fermilab and 
participating institutions. This research was supported by 
the Brazilian Conselho Nacional de Desenvolvimento Cient\'\i fico e 
Tecnol\'ogico, CONACyT (Mexico), the Israeli Academy of Sciences and 
Humanities, the U.S.-Israel Binational 
Science Foundation, and the U.S.~National Science Foundation 
and Department~of Energy. 
The Universities Research Association~operates Fermilab for 
the U.S.~Department~of Energy.
\vskip 15pt


\begin{thebibliography}{99}
\bibitem{FCNC}
E.~M. Aitala \etal (E791), \Journal{\PRL}{76}{364}{1996}.
%
\bibitem{FCNCnew}
E.~M.~Aitala \etal (E791), \Journal{\PLB}{B462}{401}{1999}; \hfil \break
David A.~Sanders, DPF\,99--Los Angeles (5--9 January 1999) hep-ex/9903067;
\hfil \break 
D.~J.~Summers \etal, XXXV Rencontre de Moriond (11--18 March 2000) 
hep-ex/0010002.
%
\bibitem{4prong}
E.~M. Aitala \etal (E791), \Journal{\PRL}{86}{3969}{2001}; \hfil \break
D.~J.~Summers, DPF2000--Columbus (9--12 August 2000) hep-ex/0011079; \hfil\break
A.~J.~Schwartz, HQ2K--Rio de Janeiro (9-12 October 2000) hep-ex/0101050;
\hfil \break
D.~A.~Sanders \etal, XXXVI Rencontre de Moriond (10-17 March 2001) 
hep-ex/0105028.
%
\bibitem{Pakvasa}
See for example: S.~Pakvasa, hep-ph/9705397; \
\Journal{\CJP}{32}{1163}{1994};\hfil\break
D.~A.~Sanders, \Journal{\MPLA}{A15}{1399}{2000}.
%
\bibitem{SCHWARTZ93} 
A.~J. Schwartz, \Journal{\MPLA}{A8}{967}{1993}.
%
\bibitem{palmer}
M.\,Banner \etal, \PLX(B122,476,1983); P.\,Bagnaia \etal, 
\PLX(B129,130,1983); \hfil \break
G.~Arnison \etal, \PLX(B122,103,1983); \PLX(B126,398,1983); 
\hfil \break  
R.~B.~Palmer, E.~A.~Paschos, N.~P.~Samios,~Ling-Lie Wang, \PRDX(D14,118,1976).
%
\bibitem{Fajfer} 
S.~Fajfer, S.~Prelov\v{s}ek, and P.~Singer, 
\Journal{\PRD}{D58}{094038}{1998}; \ hep-ph/0106333; \hfil\break
S.~Prelov\v{s}ek, ``Weak Decays of Heavy Mesons,'' hep-ph/0010106; 
\ hep-ph/0104159.
%
\bibitem{Singer}
P. Singer, \Journal{\APPB}{B30}{3861}{1999};\hfil\break
P.~Singer and D.-X.~Zhang, \Journal{\PRD}{D55}{1127}{1997}.
%
\bibitem{CPT}
G.~L\"uders, Ann.~Phys. {\bf 2} (1957) 1.
%
\bibitem{LQ}
J.~C.~Pati and A.~Salem, \PRDX(D8,1240,1973); {\bf D10} (1974) 275;
\hfil\break
Miriam Leurer, \PRDX(D49,333,1994);  {\bf D50} (1994) 536;
\hfil \break
Fermilab D0, V.~M.~Abazov \etal, Phys.~Rev. {\bf D64} (2001) 092004; 
\hfil \break
HERA H1, C.~Adloff \etal, Phys.~Lett. {\bf B523} (2001) 234.
%
\bibitem{HGB}
J.~Chakrabarti, M.~Popovic, and R.~N.~Mohapatra, \PRDX(D21,3212,1980); 
\hfil\break
Hans-Uno Bengtsson, Wei-Shu Hou, A.~Soni, and D.~H.~Stork, \PRLX(55,2762,1985). 
%
\bibitem{Ali}
A.~Ali, A.~V.~Borisov, and N.~B.~Zamorin, Eur.~Phys.~J. {\bf C21} (2001) 123.
%
\bibitem{Castro}
G.~L\'{o}pez Castro, R.~Mart\'{\i}nez, and J.~H.~Mu\~{n}oz, 
 \Journal{\PRD}{D58}{033003}{1998}.
%
\bibitem {e791spect} 
E.\,M. Aitala \etal (E791), \Journal{\EPJD}{C1}{4}{1999}; \hfil \break
L.~M.~Cremaldi, ICHEP--Dallas (6--12 August 1992), 
AIP Conf.~Proc. {\bf 272} (1993) 1058; \hfil \break
D.~J.~Summers \etal, XXVII Rencontre de Moriond (15--22 March 1992) 417, 
hep-ex/0009015;
%
\bibitem{e516}
Fermilab E516: B.~H.~Denby \etal, \PRLX(52,795,1984); \hfil\break
G.~F.~Hartner \etal, \NIMX(216,113,1983).
%
\bibitem{e691}
Fermilab E691: J.~R.~Raab \etal, \PRDX(D37,2391,1988); \hfil \break
R.J.~Morrison and M.~S.~Witherell, Ann.~Rev.~Nucl.~Part.~Sci.~{\bf 39} 
(1989) 183; \hfil \break
P.~E.~Karchin \etal, \IEEETNSX(32,612,1985); \hfil\break
B.~R.~Kumar, in
{\em Vertex Detectors,} Plenum Press, Erice (21-26 September 1986) 167.
%
\bibitem{e769}
Fermilab E769: G.~A.~Alves \etal, 
\PRDX(D49,4317,1994); \ {\bf D56} (1997) 6003; \
\PRLX(69,3147,1992); \ 
{\bf 70} (1993) 722; \ 
{\bf 72} (1994) 812; \ 
{\bf 77} (1996) 2388; \ 
2392; \ 
hep-ex/0303027; \hfil\break
D.~J.~Summers \etal, XXIII Rencontre de Moriond (13-19 March 1988) 451,
   hep-ex/0011020; \hfil\break
D.~Errede \etal, \NIMX(A309,386,1991);
\IEEETNSX(36,106,1989); \hfil\break
C.~Gay and S.~Bracker, {\it ibid.,} {\bf 34} (1987) 870; \
R.~Vignoni \etal, {\it ibid.,} {\bf 34} (1987) 756; \
O.~Calvo and G.~Kraft, {\it ibid.,} {\bf 36} (1989) 770; \,
C.\,Stoughton and D.\,Summers, \CIP(6,371,1992); \hfil \break
P.\,Karchin \etal, Nucl.\,Phys.\,Proc.\,Suppl. {\bf 7B} (1989) 60;
R.\,Jedicke, Phys.\,Canada {\bf 45} (1989) 105.  
%
\bibitem{mix}
E.M.~Aitala \etal (E791), \PRLX(77,2384,1996); \
                           \PRDX(D57,13,1998); \hfil \break
                           \PRLX(83,32,1999); \
L.~Cremaldi, Nucl.~Phys.~Proc.~Suppl.~{\bf 55A} (1997) 221; \hfil \break
A.~Napier \etal, St.~Petersburg 1994, AIP Conf.~Proc. {\bf 338} (1995) 243;
\hfil \break 
Jun-ichi Tanaka (KEK Belle), hep-ex/0104053; AIP Conf.~Proc. {\bf 618}
(2002) 293;  \hfil \break
K.~Abe \etal. (KEK Belle), \PRLX(88,162001,2002); \hfil \break 
B.~Aubert \etal (SLAC BaBar), hep-ex/0304007; hep-ex/0306003;  \hfil \break
J.~M.~Link \etal (Fermilab FOCUS), \PLX(B485,62,2000); \hfil \break
R.~Godang \etal (CLEO), \PRLX(84,5038,2000); \hfil \break
J.~C.~Anjos \etal (Fermilab E691), \PRLX(60,1239,1988).
%
\bibitem{CP}
E.~M.~Aitala \etal (E791), \PLX(B403,377,1997); \ \PLX(B421,405,1998);
\hfil \break
J.~Bartelt \etal (CLEO), \PRDX(D52,4860,1995); \hfil \break
P.~L.~Frabetti \etal (Fermilab E687), \PRDX(D50,2953,1994); \hfil \break
J.~C.~Anjos \etal (Fermilab E691), \PRDX(D44,3371,1991).
%
\bibitem{pentaquark}
E.~M.~Aitala \etal (E791), \PRLX(81,44,1998); \ \PLX(B448,303,1999).
%
\bibitem{cabibbo}
E.~M.~Aitala \etal (E791), \PLX(B404,187,1997).
%
\bibitem{Ds_life}
E.M.~Aitala \etal (E791), \PLX(B445,449,1999).
%
\bibitem{Dpi}
E.~M.~Aitala \etal (E791), \PLX(B403,185,1997); \hfil \break  
K.~Gounder and L.~Cremaldi, DPF\,96--Minneapolis (11--15 August 1996) 868,
hep-ex/0008061. \hfil\break
CDF has used $B$-$\pi$ production correlations to do
$B^0 \, \overline{B}^{\,0}$ tagging and
measure {\it CP} violation in
$B^0 \rightarrow J/\psi \, \, K^0_{\hbox{s}}$ decays: \
F.~Abe \etal, \PRLX(81,5513,1998); \
J.~Kroll, hep-ex/9908062;
G.~Bauer, hep-ex/9904017; hep-ex/9908055; hep-ex/0011076;
NIM {\bf A253} (1987) 179.
%
\bibitem{Sigma}
E.~M.~Aitala \etal (Fermilab E791), \PLX(B379,292,1996); \hfil \break  
J.~C.~Anjos \etal (Fermilab E691), \PRLX(62,1721,1989).
%
\bibitem{Dplus_prod}
E.\,M.~Aitala \etal (E791), \Journal{\PLB}{B371}{157}{1996}.
%
\bibitem{Lambda_c_prod}
E.M.~Aitala \etal (E791), \PLX(B495,42,2000).
%
\bibitem{Ds_prod}
E.M.~Aitala \etal (E791), \PLX(B411,230,1998). 
%
\bibitem{D0_cross}
E.M.~Aitala \etal (E791), \PLX(B462,225,2000).
%
\bibitem{dstar}
E.M.~Aitala \etal (E791), \PLX(B539,218,2002).
%
\bibitem{Lambda_c}
E.M.~Aitala \etal (E791), \PLX(B471,449,2000); \hfil \break
M.~V.~Purohit, ICHEP2000--Osaka (27 July -- 2 August 2000) hep-ex/0010038;
\hfil \break
Hyperons, Charm, and Beauty Hadrons\,98--Genoa,
Nucl.\ Phys.\ Proc.\ Suppl.\ {\bf 75B} (1999) 208. 
%
\bibitem{KKpipi}
E.M.~Aitala \etal (E791), \PLX(B423,185,1998).
%
\bibitem{Sokoloff}
E.M.~Aitala \etal (E791), \PRDX(D64,112003,2001); 
%
\bibitem{Gobel}
Carla Gobel, HQ2K--Rio de Janeiro (9-12 October 2000) hep-ex/0012009; 
\hfil \break
Carla Gobel, Hadron 2001--Protvino (25 Aug -- 1 Sep 2001) hep-ex/0110052; 
\hfil \break
E.M.~Aitala \etal (E791), Phys.~Rev.~Lett. {\bf 89} (2002) 121801. 
%
\bibitem{Ds3pi}
E.M.~Aitala \etal (E791), \PRLX(86,765,2001); \hfil \break
J. C. Anjos \etal (Fermilab E691), \PRLX(62,125,1989).
%
\bibitem{D3pi}
E.M.~Aitala \etal (E791), \PRLX(86,770,2001); \hfil \break
Ignacio Bediaga, QCD\,00--Montpellier (6--13 July 2000),
{Nucl.\,Phys.\,Proc.\,Suppl.} {\bf 96} (2001) 225; \hfil \break
J.~M.~de Miranda, Comments Nucl.\,Part.\,Phys. {\bf 2} (2002) A362; \hfil\break
P.~Estabrooks, \PRDX(D19,2678,1979); \hfil \break 
R.~Gatto, G.~Nardulli, A.~D.~Polosa, and N.~A.~Tornqvist, \PLX(B494,168,2000).
%
\bibitem{hyperon}
E.M.~Aitala \etal (E791), \PLX(B496,9,2000).
%
\bibitem{Valence}
E.M.~Aitala \etal (E791), \PRLX(86,4768,2001); \hfil \break
L.~Frankfurt, G.~A.~Miller, and M.~Strikman, Phys.~Rev. {\bf D65} (2002) 094015.
%
\bibitem{Color}
E.M.~Aitala \etal (E791), \PRLX(86,4773,2001).
%
\bibitem{da791}
S.~Amato \etal, \NIMX(A324,535,1993); \hfill \break
Steve Bracker and Sten Hansen,  hep-ex/0210034; \hfil \break 
S.~Hansen \etal, \IEEETNSX(34,1003,1987); \hfill \break
A.~E.~Baumbaugh \etal, {\it ibid.} {\bf 33} (1986) 903; \
K.~L.~Knickerbocker \etal, {\it ibid.} {\bf 34} (1987) 245. \hfill \break
High speed systems similar to the E791 DA have subsequently been built by 
WASA at the Svedberg Accelerator Lab in Sweden,
by E864 at Brookhaven, and by HYPERCP at Fermilab. \hfill \break 
L.~Gustafsson \etal (WASA), \IEEETNSX(41,1155,1994); \hfill \break
John Lajoie, ``The BNL E-864 Data Acquisition System: A High Speed,
Parallel DA System for Particle Physics Experiments," IEEE RT\,95--East Lansing
(22-25 May 1995); \hfill \break
Y.~C.~Chen \etal (Fermilab HYPERCP), \NIMX(A455,424,2000).
%
\bibitem{farm791}
S.~Bracker \etal, \IEEETNSX(43,2457,1996); \hfill \break
D.~Summers \etal, DPF\,96--Minneapolis (11--15 August 1996) 1385,
hep-ex/0007003; \hfill \break
F.~Rinaldo and S.~Wolbers, \CIP(7,184,1993); \
Jeffery A.~Appel, 9th Mexican School on Particles and Fields--Puebla 
(9-19 August 2000) hep-ex/0010074.
%
\bibitem{Bartlett}
D. Bartlett \etal, \NIMX(A260,55,1987).
%
\bibitem{SLIC}
E.~M.~Aitala \etal (E791), \PRLX(80,1393,1998); \hfill \break
V.~K. Bharadwaj \etal, \NIMX(155,411,1978); {\bf A228} (1985) 283; \hfill \break
D.~J. Summers,  {\it ibid.} {\bf A228} (1985) 290; \
J.~A.~Appel \etal, {\it ibid.} {\bf A243} (1986) 361.
%
\bibitem{muon}
E.~M.~Aitala \etal (E791), \PLX(B450,294,1999); \
                           \PLX(B397,325,1997).
%
\bibitem{Chong}
E.~M.~Aitala \etal (E791), \PLX(B440,435,1998).
%
\bibitem {MC} 
 H.-U.~Bengtsson and T.~Sj\"{o}strand, \Journal{\CPC}{82}{74}{1994}; \
T.~Sj\"{o}strand, Pythia 5.7 and Jetset 7.4: Physics and Manual, 
LU-TP-95-20, CERN-TH-7112-93-REV, hep-ph/9508391.
%
\bibitem {PDG98}
Particle Data Group, C.~Caso \etal, \Journal{\EPJC}{C3}{1}{1998}.
%
\bibitem{Feldman}
G.~J.~Feldman and R.~D.~Cousins, \Journal{\PRD}{D57}{3873}{1998}.
%
\bibitem{Cousins}
R.~D. Cousins and V.~L. Highland, \Journal{\NIMA}{A320}{331}{1992}.
%
\bibitem{E687}
Fermilab E687 Collaboration, P.~L.~Frabetti \etal, \PLX(B398,239,1997); 
\hfil\break
Jianwei Cao, Ph.\,D. Dissertation, Vanderbilt University, 1997.

%
\bibitem{E653}
Fermilab E653 Collaboration, K.~Kodama \etal, \PLX(B345,85,1995).
%
\bibitem{E771}
CERN BEATRICE (WA92) Collaboration, M.~Adamovich \etal, \PLX(B408,469,1997); 
\hfil\break
Fermilab E771 Collaboration, T.~Alexopoulos \etal, \PRLX(77,2380,1996).
%
\bibitem{E789}
Fermilab E789 Collaboration, D.~Pripstein \etal, \PRDX(D61,032005,2000).
%
\bibitem {PDG2000}
Particle Data Group, D.~E.~Groom \etal, \Journal{\EPJC}{C15}{1}{2000}.
%
\bibitem{CLEO}
CLEO Collaboration, A.~Freyberger \etal, \PRLX(76,3065,1996).
%
\bibitem{Sheldon}
J. M. Link \etal (Fermilab FOCUS), hep-ex/0306049; \hfil \break
Paul D.~Sheldon, Heavy Flavors 8--Southampton (25--29 July 1999)
hep-ex/9912016; \hfil \break
Fermilab E687 Collaboration, P.~L.~Frabetti \etal, \NIMX(A320,519,1992).  
%
\bibitem{aitala}
Eric Aitala and David Sanders, ``Search for Rare and Forbidden Decays of
4--Particle Decay Modes of $D^0$," E791 Offline Document 434, 42 pp., 2001;
\hfil \break
Eric Matthew Aitala (Fermilab E791), ``A Study of Atomic Number Dependence of
Charm D Meson Hadroproduction," Master's Thesis, University of Mississippi,
1993.
%
\bibitem{appel}
J.\,A.~Appel, Ann.\,Rev.\,Nucl.\,Part.\,Sci.\ {\bf 42} (1992) 367; \hfil \break
J.\,A.~Appel, SLAC Beam Line {\bf 28N1} (1998) 17.
%
\bibitem{bracker}
B.~M.~Lasker, S.~B.~Bracker, and W.~E.~Kunkel,
Publ.~Astron.~Soc.~Pac. {\bf85} (1973) 109.
%
\bibitem{cremaldi}
L.~M.~Cremaldi, HQ93 Frascati, Frascati Phys.~Ser. {\bf 1} (1993) 245; 
\hfil \break 
L.~M.~Cremaldi, 1987 Stanford Heavy Flavors, 
Annals N.~Y.~Acad.~Sci. {\bf 535} (1988) 433.
%
\bibitem{nguyen}
A.~Nguyen \etal, St.~Petersburg 1994, AIP Conf.~Proc. {\bf 338} (1995) 777; 
%
\bibitem{purohit}
M.~V.~Purohit, HQ96 St.~Goar, Frascati Phys.~Ser. {\bf 7} (1997) 269; 
\hfil \break
M.~V.~Purohit, HQ94 Charlottesville, Frascati Phys.~Ser. {\bf 3} (1994) 97; 
\hfil \break
M.~V.~Purohit, HQ93 Frascati, Frascati Phys.~Ser. {\bf 1} (1993) 63.
%
\bibitem{sanders}
David Sanders, ``Search for Rare and Forbidden Dilepton Decays of $D^+$,
$D_s^+$, and $D^0$, E791 Offline Document 393, 77 pp., 1999; \
M.\,Arneodo... D.\,Sanders \etal, \PLX(B332,195,1994); \hfil \break
D.~B.~Cline, D.~A.~Sanders, and W.~Hong,
Astrophys.~J. {\bf 486} (1997) 169; \hfil \break
D.~A.~Sanders {\it et al.,} IEEE Trans.~Nucl.~Sci. {\bf 49} (2002) 1834; 
\hfil \break
D.~A.~Sanders {\it et al.,} CHEP\,03, La Jolla, arXiv.org/pdf/physics/0306037. 
%
\bibitem{slaughter}
A.~J.~Slaughter, ``Rare and Forbidden Decays of Charm and Beauty, Mixing
in the Charm Sector," 
Proceedings, 16th Physics in Collision, Mexico City (19-21 June 1996)
189; \hfill \break
J.~Slaughter, CIPANP\,97, Big Sky, Montana, AIP Conf.~Proc. {\bf 412} (1997)
648.
%
\bibitem{stanton}
N.~R.~Stanton {\it et al.,}  Phys.~Rev.~Lett. {\bf 42} (1979) 346; \hfil \break
N.~R.~Stanton (E653), 1991 San Miniato Heavy Flavors,
Nucl.~Phys.~Proc.~Suppl. {\bf 27} (1992) 201; \hfil \break
K.~Kodama... N.~R.~Stanton {\it et al.} (Fermilab DONUT), 
Phys.~Lett. {\bf B504} (2001) 218.
%
\bibitem{witchey}
Nicholas James Witchey (Fermilab E791), ``Search for Flavor Changing Neutral
Current Decays of Charm Mesons," Ph.D. Dissertation, 
The Ohio State University, 1996.

\end{thebibliography}
\end{document}